\documentclass[journal=jacsat,manuscript=article]{achemso}
\setkeys{acs}{articletitle = true}

\usepackage[version=3]{mhchem} 
\usepackage{amsmath}  
\usepackage{amsfonts} 
\usepackage{subfig}
\usepackage{graphicx} 
\usepackage{booktabs}
\usepackage{threeparttable}
\usepackage[font=footnotesize]{caption}
\usepackage{verbatim}
\usepackage{xcolor}
\usepackage{soul}

\author{Benjamin Rudshteyn\textsuperscript{\#}}
\author{Dilek Coskun\textsuperscript{\#}}
\author{John L. Weber}
\affiliation{Department of Chemistry, Columbia University, 3000 Broadway, New York, NY, 10027}
\author{Evan J. Arthur}
\affiliation{Schrodinger Inc., 120 West 45th Street, New York, NY, 10036}
\author{Shiwei Zhang}
\affiliation{Center for Computational Quantum Physics, Flatiron Institute, 162 5th Avenue, New York, NY 10010}
\alsoaffiliation{Department of Physics, College of William and Mary, Williamsburg, VA 23187}
\author{David R. Reichman}
\author{Richard A. Friesner}
\author{James Shee\textsuperscript{\%,}}
\email{js4564@columbia.edu}
\affiliation{Department of Chemistry, Columbia University, 3000 Broadway, New York, NY, 10027}

\title{Predicting Ligand-Dissociation Energies of 3$d$ Coordination Complexes with Auxiliary-Field Quantum Monte Carlo} 

\begin{document}
\textsuperscript{\#} These authors made equal contributions.

\textsuperscript{\%} Current address: Department of Chemistry, University of California, Berkeley, California 94720


\begin{abstract}
Transition metal complexes are ubiquitous in biology and chemical catalysis, yet they remain difficult to accurately describe with \emph{ab initio} methods due to the presence of a large degree of dynamic electron correlation, and, in some cases, strong static correlation which results from a manifold of low-lying states.  Progress has been hindered by a scarcity of high quality gas-phase experimental data, while exact \emph{ab initio} predictions are usually computationally unaffordable due to the large size of the relevant complexes.  In this work, we present a data set of 34 tetrahedral, square planar, and octahedral 3$d$ metal-containing complexes with gas-phase ligand-dissociation energies that have reported uncertainties of $\leq$ 2 kcal/mol. We perform all-electron phaseless auxiliary-field quantum Monte Carlo (ph-AFQMC) calculations utilizing multi-determinant trial wavefunctions selected by a blackbox procedure.  We compare the results with those from density functional theory (DFT) with the B3LYP, B97, M06, PBE0, $\omega$B97X-V, and DSD-PBEP86/2013  functionals, and a localized orbital variant of coupled cluster theory with single, double, and perturbative triple excitations (DLPNO-CCSD(T)). We find mean averaged errors of 1.09 $\pm$ 0.28 kcal/mol for our best ph-AFQMC method, vs 2.89 kcal/mol for DLPNO-CCSD(T) and 1.57 - 3.87 kcal/mol for DFT.
We find maximum errors of 2.96 $\pm$ 1.71 kcal/mol for our best ph-AFQMC method, vs 9.15 kcal/mol for DLPNO-CCSD(T) and 5.98 - 13.69 kcal/mol for DFT. The reasonable performance of a number of DFT functionals is in stark contrast to the much poorer accuracy previously demonstrated for diatomic species, suggesting a moderation in electron correlation due to ligand coordination. However, the unpredictably large errors for a small subset of cases with both DFT and DLPNO-CCSD(T) methods
leave cause for concern, especially in light of the unreliability of common multi-reference indicators. In contrast, the robust and, in principle, systematically improvable results of ph-AFQMC for these realistic complexes establish the method as a useful tool for elucidating the electronic structure of transition metal-containing complexes and predicting their gas-phase properties.

\end{abstract}

\section{Introduction}

The unique electronic structure of transition metals enables a rich variety of chemical reactivity, harnessed in systems
ranging from those found in the fields of chemical catalysis,\cite{prier2013visible}
biology\cite{trautwein1997bioinorganic} and materials science\cite{khomskii2014transition}.  The presence of multiple quantum states within an accessible energy range allows for reaction mechanisms involving sequential redox events and subtle transformations between spin-states, e.g. in clusters of Mn atoms in Photosystem II (PSII) or Fe and Mo atoms in nitrogenases\cite{Siegbahn_2017,Raugei_Seefeldt_Hoffman_2018,askerka2016oec,siegbahn2017nucleophilic}.  Furthermore, the coordination of small molecules to single metal ions is an important motif in drug design\cite{riccardi2018metal}, and the
correlations exhibited in the copper oxide layers of cuprate materials play a central role in the phenomenon of high-temperature superconductivity\cite{lee2006doping,dagotto1994correlated}. 

\emph{Ab initio} modeling has the potential to yield essential insights into these transition metal systems.  However, exact methods scale exponentially with system size and are thus only applicable to small molecules.  
Many groups have used density functional theory (DFT) to examine the electronic structure and reaction mechanisms of coordinated transition metal complexes, including the active sites of PSII\cite{askerka2016oec,siegbahn2017nucleophilic} and 
cytochrome P450,\cite{Friesner_2011,Jerome_Hughes_Friesner_2016}  catalysts for water oxidation,\cite{rudshteyn2018water} CO$_2$ reduction, \cite{agarwal2012mechanisms} and sensitizers for optical upconversion.\cite{han2016metallophthalocyanines}  
However, there are a number of uncertainties which may 
cast doubt upon
their conclusions, chief among them possible errors due to electron self-interaction  and strong correlation.  Furthermore, as the majority of parameterized density functionals and dielectric continuum solvation models have been trained on organic compounds (e.g. the $\omega$B97X-V \cite{mardirossian2014omegab97xv} and $\omega$B97M-V\cite{mardirossian2016omegab97mv} functionals  and the SMD solvation model\cite{marenich2009universal}), 
it is reasonable to suspect the accuracy of the resulting predictions in the domain of transition metal chemistry.

The pronounced lack of reliable and precise gas-phase experimental data for realistic transition metal systems, as illustrated by recent theoretical benchmarking studies, exacerbates these issues.\cite{bross2013explicitly,manivasagam2015pseudopotential,jiang2012comparative,zhang2013tests,moltved2018chemical,jiang2011toward,carlson2014multiconfiguration,bao2017predicting,baobimetallic,sharkas2017multiconfiguration,tran2018spin}  
This scarcity of experimental measurements is in stark contrast to the large amount of reliable experimental values for organic molecules, which has enabled very accurate parameterizations of DFT functionals and a thorough validation of methods such as CCSD(T), which can readily achieve $\sim$1 kcal/mol accuracy for typical organic molecules \cite{bartlett2007coupled}.

The accuracy of CC methods, most frequently CCSD(T), is often assumed to carry over to transition metal systems, as evidenced by a number of studies that have attempted to draw conclusions about the accuracy of DFT by comparing against reference CC values.\cite{quintal2006benchmark,steinmetz2013benchmark,kang2012accurate,dohm2018comprehensive,chan2019assessment}. 
However, the reliability of CC methods for transition metal systems, even when multireference effects are approximated, has been the subject of vigorous debate, as illustrated by recent studies on transition metal diatomic-ligand systems. \cite{jiang2011toward,Xu_Truhlar_JCTC_2015,cheng2017bond,fang2017prediction,aoto2017arrive,shee2019achieving,williams2019direct}.  
de Oliveira-Filho and co-workers found that even multireference CCSD(T) could not predict the bond dissociation energies (BDEs) for some diatomics accurately with respect to experimental measurements.
A recent study by Head-Gordon and co-workers found that high levels of CC, up to CCSDTQ, are required for chemical accuracy against an exact method known as Adaptive Sampling Configuration Interaction (ASCI) results, albeit in a small basis set.\cite{hait2019levels}  Wilson and co-workers collected a set of 225 heats of formation for compounds with first row transition metal atoms.\cite{jiang2011toward} They found good performance for their composite CC scheme vs. a subset of experimental data with small uncertainties, but the mean absolute error (MAE) of around 3 kcal/mol may be insufficient for many chemical applications. 
Reiher and co-workers considered transition metal ligand-dissociation energies of very large molecules and showed that a localized variant of  CCSD(T) 
utilizing
domain\textcolor{purple}{--}based pair natural orbitals (DLPNO-CCSD(T))\cite{guo2018communication,riplinger2016sparse} resulted in pronounced errors, e.g. $\sim$ 9.3 kcal/mol for the cleavage of a Cu complex.\cite{husch2018calculation}

An alternative benchmarking approach involves filtering out strongly correlated cases with multireference diagnostics, and benchmarking DFT against CC methods only for the single-reference subset of molecules.  
Hansen, Checinski, and co-workers developed the MOR41 test set of organometallic reactions of medium-large size.  They removed 
open-shell, multi-reference cases (with, e.g., FOD and T1 diagnostics).  
Recently, the properties of a set of transition metal atoms and oxide diatomics, in which strongly multi-reference cases were removed, were predicted by a large number of $\emph{ab initio}$ methods.\cite{williams2019direct} 
In our view, this strategy is less than ideal not only because a large subset of relevant chemistry is excluded, but moreover because the utility of affordable multi-reference indicators has increasingly been called into question.  Indeed, studies have found mixed success for different kinds of multireference diagnostics\cite{Xu_Truhlar_JCTC_2015,cheng2017bond,fang2017prediction,aoto2017arrive,momeni2015td}  making it hard to judge $\emph{a priori}$ when single-reference methods would be appropriate.

In this work, we assemble a test set of gas-phase ligand-dissociation measurements with low reported experimental uncertainties.  On this set we use auxiliary field quantum Monte Carlo with the phaseless constraint (ph-AFQMC),\cite{zhang2003quantum,al2006auxiliary} accelerated by a correlated sampling technique\cite{shee2017chemical} and our implementation on graphical processing units\cite{shee2018gpu}.  We have shown that this method yields robust accuracy for the ionization potential of transition metal atoms\cite{shee2018gpu} and the dissociation energy of transition metal-containing diatomics\cite{shee2019achieving}. 
The present study marks a large step forward, to more relevant transition metal-containing systems. We demonstrate that ph-AFQMC with correlated sampling yields accurate BDE predictions for various tetrahedral, square planar, and octahedral complexes containing first row transition metal atoms and ligands including dihydrogen, chloride, dinitrogen, aqua, ammonia, carbonyl, and formaldehyde.  We then validate the performance of a representative set of DFT functionals and the DLPNO-CCSD(T) method.  
Consistent with our expectation, we find that
single-reference methods such as DFT and the CC hierarchy 
perform better for coordinated metal compounds compared to the case of diatomic dissociation (as ligand coordination can lower the degree of degeneracy of the metal atomic $d$ orbitals). However,
we demonstrate that ph-AFQMC still produces a significant improvement in terms of MAE and maximum error (MaxE). 

Our results show that ph-AFQMC can consistently produce benchmark-quality results, and with a computational cost which scales as a low polynomial with system size (excluding the cost of obtaining the CASSCF trial wavefunctions).  This method will extend accurate reference datasets for future benchmarking studies of approximate methods such as DFT and accurate classical potentials for transition metal ions.  In addition, the level of accuracy of the widely-employed quantum-chemical methods included in this study provides a sense of the accuracy to be expected for calculations on similar 4- and 6- coordinated 3$d$ metal complexes that are ubiquitous in fields such as biology and catalysis.
\section{Selection of Experimental Data} 

\begin{figure}[H]
    \centering
    \includegraphics[width=14cm]{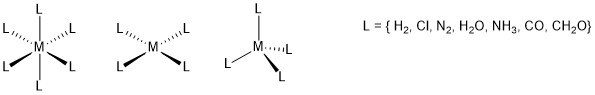} 
    \caption{ The types 
    of transition metal compounds studied. M can be any 3$d$ transition metal from Ti to Cu.} 
    \label{fig:coordination_compounds_final}
\end{figure}

We selected gas-phase experimental BDE data with less than or equal to 2.0 kcal/mol uncertainty from the recommended values in the handbook compiled by Luo \cite{luo2007comprehensive}. Most of the measurements can also be found in the work by Rodgers and Armentrout.\cite{rodgers2016cationic}  For TiCl$_4$, Hildenbrand's updated experimental measurement has been used.\cite{hildenbrand2009low} The average uncertainty for the molecules included in the present test set is 1.03 kcal/mol. Most of the measurements were performed with the threshold collision-induced dissociation technique except for  [Ni(H$_2$O)$_6$]$^{2+}$, TiCl$_4$, CrCO$_5$H$_2$ and V(H$_2$O)(H$_2$)$_3$ which were measured with blackbody infrared radiative dissociation, effusion beam mass spectroscopy, transient infrared spectroscopy for kinetic analysis and temperature-dependent equilibrium, respectively.  The latter technique was used for all other H$_2$ complexes as well.  The selected compounds are depicted schematically in Fig. \ref{fig:coordination_compounds_final}. These experimental data are mostly extrapolated to 0\,K, and can therefore be directly compared with quantum-chemical calculations. The two exceptions are TiCl$_4$ and CrCO$_5$H$_2$, which are measured at 298 K. All the metal complexes have +1 net charge, except for [Ni(H$_2$O)$_6$]$^{2+}$, TiCl$_4$, and CrCO$_5$H$_2$. The full list of reactions is given in the Supporting Information (SI).

\section{Computational Details} 

The geometries, reorganization energies (\textit{vide infra}), and enthalpic corrections (just the zero-point energy (ZPE) for cases where the 0 K extrapolated experiment is available, as discussed above) were obtained with DFT calculations with the B3LYP functional \cite{becke1993b3lyp,vosko1980accurate,lee1988development} and cc-pVTZ-dkh\cite{dunning1989gaussianbasis,woon1993gaussianbasis,jong2001parallel,balabanov2005systematically} basis set using the ORCA program package.\cite{neese2012orca} Details regarding occasional small imaginary frequencies and integration grids are given in Section IV of the SI.
The DLPNO-CCSD(T) calculations were also done with ORCA using ``TightPNO" localization parameters and the cc-pV$x$Z-dkh basis sets, $x$=T,Q, and are extrapolated to the complete basis set limit using the procedure built into ORCA,\cite{neese2012orca} as discussed in the SI.
The DKH2 relativistic correction was used for all DFT and CC calculations.\cite{pantazis2008all}   

Integrals for AFQMC were obtained with PySCF\cite{sun2018pyscf}. The exact-two-component (x2c) relativistic Hamiltonian\cite{liu2009exact} was used in place of DKH2. As in our previous work,\cite{shee2017chemical,shee2018gpu, shee2019achieving, shee2019singlettriplet} the imaginary time step for the AFQMC propagation, utilizing single precision floating point arithmetic, was 0.005 Ha$^{-1}$. The walker orthonormalization, population control, and local energy measurements occurred every 2, 20, and 20 steps, respectively. We utilized a modified Cholesky decomposition of the electron repulsion integrals with a cutoff of 10$^{-5}$. Walkers were initialized with the RHF/ROHF determinant.

\begin{figure}[H]
    \centering
    \includegraphics[width=7cm]{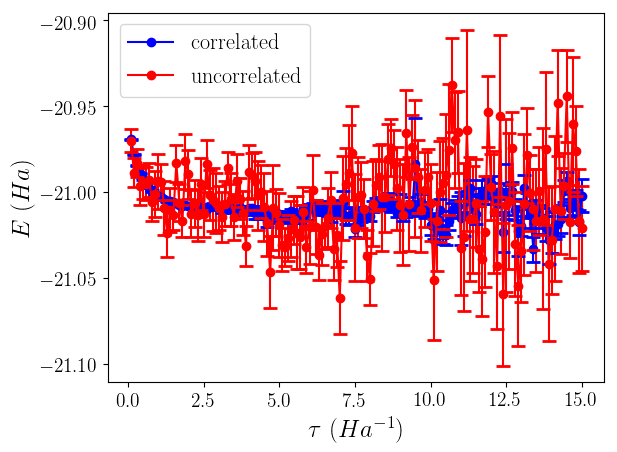} 
    \caption{Correlated sampling ph-AFQMC calculations  using Summit GPU’s; statistical errors from correlated and uncorrelated sampling approaches are compared for the Cu-H$_{2}$ bond dissociation energy of the [Cu(H$_{2}$)$_{4}$]$^+$ molecule.} 
    \label{fig:Cu_H24_BDE}
\end{figure}

The correlated sampling approach\cite{shee2017chemical} can converge energy differences between similar states by employing a shared set of auxiliary fields for a short projection time, providing accurate results with smaller statistical errors vs uncorrelated AFQMC (the latter would need to run longer projections to reach the same statistical accuracy).  This approach performs most efficiently when the ligand being removed is small, as indicated by our previous work in which the reduction in statistical error vs the uncorrelated approach was several times larger for MnH than for MnCl.\cite{shee2019achieving} Similar behavior is found  for the transition metal complex systems studied here, as shown in Fig. \ref{fig:Cu_H24_BDE} for [Cu(H$_{2}$)$_{4}$]$^+$.  In fact, correlated sampling may work better for these complexes than it did for the diatomics since $<<$ 50\% of the system is being changed. Finally, we note that correlated sampling also can improve the accuracy of the predicted results in certain situations. \cite{shee2018gpu,shee2019achieving}

In the context of computing BDEs, our AFQMC calculations used correlated sampling for the difference in energy between the original coordination compound (M-L) and the species missing a ligand (M), i.e. the same geometry but with ghost basis functions centered around the positions of the missing nuclei that comprise the ligand. 
If the difference in energies was not converged before 15 Ha$^{-1}$, uncorrelated, separate AFQMC calculations are performed for the optimized structures of both states without ghost basis functions, using a population control scheme in which walkers with large weights are duplicated while
those with small weights are randomly destroyed for the optimized structures of both states without ghost basis functions\cite{motta2018ab}. The isolated ligand (L) was also treated with the population control approach.

\begin{figure}[H]
    \centering
    \includegraphics[width=7cm]{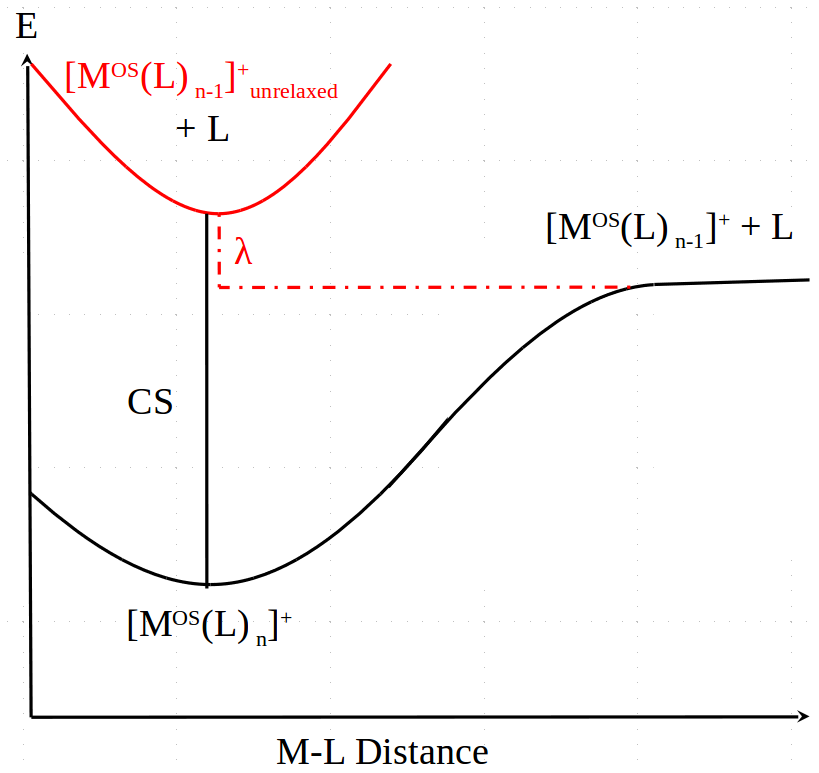} 
    \caption{Schematic of BDE calculations performed in this work.  OS abbreviates oxidation state, CS indicates the energy measured by the correlated sampling approach.} 
    \label{fig:PESplot}
\end{figure}

The BDE is given as follows:
\begin{equation}
    BDE = (H(M) - H(M-L)) + H(L) - \lambda,
\end{equation}

where $H$ are enthalpies including the zero-point corrections and the nuclear repulsion energy.  The reorganization energy, $\lambda$, is defined as the difference in energy between the product (complex with the ligand dissociated) in its optimal geometry and in the reactant geometry, optimized with the ligand, but with the ligand atoms deleted.  $\lambda$ is computed via DFT.
The calculation of BDEs is illustrated in Fig. \ref{fig:PESplot}.

To give a sense of the required computational cost, a correlated sampling ph-AFQMC calculation for [Fe(N$_2$)$_4$]$^+$ 
took about 267 node hours on Summit, using a truncated CASSCF trial wave function containing 1195 determinants. This reflects the use of 20 repeats (i.e. independent trajectories with different random number seeds), each using 20 nodes with 6 GPU's each (each repeat ran for about 42 minutes). 

The complete basis set limit for the ph-AFQMC calculations was estimated by extrapolation using DLPNO-CCSD(T) values with the cc-pV$x$Z-dkh basis sets, $x$=T,Q, using exponential and $\frac{1}{x^3}$ forms for the mean-field (i.e. UHF) and correlation energies, respectively, as in our previous work\cite{shee2019achieving}. We used the equivalent cc-pV$x$Z\textbackslash C auxiliary basis sets for the DLPNO approximations.
If the ph-AFQMC correlation energy with cc-pVTZ-dkh is significantly different from DLPNO-CCSD(T), or if comparison of the extrapolated value with experiment indicates a potential problem (our target accuracy is $<$3 kcal/mol, which has been referred to as ``transition metal chemical accuracy"\cite{deyonker2007quantitative}), then full extrapolation within ph-AFQMC is performed utilizing both cc-pVTZ-dkh and cc-pVQZ-dkh basis sets (for dihydrogen or chloro compounds).
In some cases, we instead extrapolate with a UHF trial-based ph-AFQMC procedure, which seems to be a good compromise between speed and accuracy (see Tables S4 and S5 for details).   

Apart from the basis set extrapolations, the ph-AFQMC calculations utilized CASSCF trial wavefunctions. The size of the CASSCF trial wavefunction for the metal-containing species was automatically selected via the AVAS procedure where only those B3LYP ROKS orbitals that overlap significantly with the 3d and/or 4d atomic orbitals (from the minimal atomic basis set called "MINAO" as used by Knizia\cite{knizia2013intrinsic} or from the Atomic Natural Orbital (ANO-RCC) basis set) of the metal were included (as noted in the SI)\cite{sayfutyarova2017automated}. The single numerical overlap threshold parameter was used to generate sequentially 
larger active spaces to determine what active space size is needed to reach chemical accuracy.
The active space for the ligand was selected by either using the valence set of electrons and orbitals or using a large number for electrons and orbitals to ensure convergence. Typically $>$98\% of the weight of the CI coefficients was retained. The active spaces were selected so that the active space for the reactant and product metal species were similar (either the same or off by 1 orbital and 2 electrons), which often requires the same AVAS threshold.

We compare ph-AFQMC with the B3LYP, M06 \cite{Zhao2008m06}, and PBE0\cite{adamo1999toward} functionals since they are arguably the most popular, and B97 since
this functional performed
the best in our previous study.\cite{shee2019achieving}  
To explore the performance of range-correction and the non-local correlation approach, we include the $\omega$B97X-V functional.\cite{mardirossian2014omegab97xv}  
We also consider the double hybrid functional, DSD-PBEP86.  It is available in ORCA, and has been shown to perform very well,\cite{kozuch2011dsd,kozuch2013spin,goerigk2017look,santra2019minimally} accelerated by the resolution of identity (RI) approximation on the MP2 part.
In this study,  we used the "DSD-PBEP86/2013" functional, which has slightly different parameters than DSD-PBEP86, 
but
refer to it as DSD-PBEP86 throughout the paper.


Since analytical gradients have not yet been implemented in ORCA for all of the functionals in this study, we decided to use B3LYP optimized geometries and performed single-point energy calculations.  Grid and density-initialization choices are described in Section IV of the SI.


For all DFT and HF (the latter is used as a reference wavefunction for DLPNO-CCSD(T)) calculations, we found it essential to perform a stability analysis to ensure that the lowest energy SCF solution was obtained.

\section{Results and Discussion}

The deviations of the computed BDEs from experiment are presented in Figs\textcolor{purple}{.} \ref{fig:LigandPlot_H2} to \ref{fig:LigandPlot_MISC}. Values of the BDEs are given explicitly in Tables S1 and S6. 
 Tables \ref{table:Dihydrogentable} through \ref{table:MAEtable} show statistical metrics including Mean Signed Error (MSE), MAE, and  MaxE for each ligand type, and ultimately for the entire test set. 

\subsection{Dihydrogen Complexes}
\begin{figure}[H]
    \centering
    \includegraphics[width=14cm]{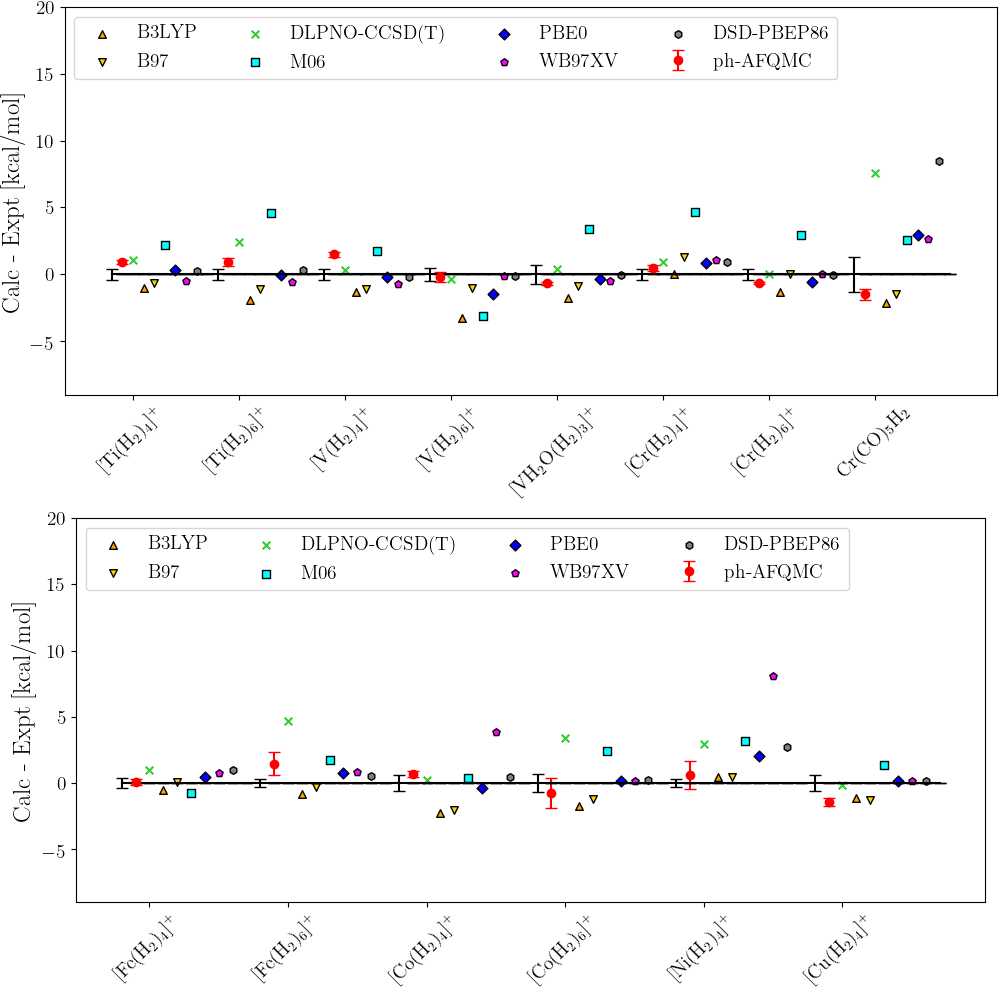} 
    \caption{Deviations [kcal/mol] of computational methods for the dihydrogen set of bond dissociation reactions where the H$_2$ that leaves is given at the end of the formula.} 
    \label{fig:LigandPlot_H2}
\end{figure}

In general, as shown in Fig. \ref{fig:LigandPlot_H2} and Table \ref{table:Dihydrogentable}, the performance of ph-AFQMC is excellent for dihydrogen complexes (where the dihydrogen is the ligand being removed), including 
[Ti(H$_2$)$_4$]$^+$, 
[Cu(H$_2$)$_4$]$^+$,
[V(H$_2$)$_4$]$^+$,	
[V(H$_2$)$_6$]$^+$,
[Co(H$_2$)$_4$]$^+$,
[Ni(H$_2$)$_4$]$^+$,
[Ti(H$_2$)$_6$]$^+$,
[Co(H$_2$)$_6$]$^+$,
[Fe(H$_2$)$_6$]$^+$,
[Fe(H$_2$)$_4$]$^+$,
[Cr(CO)$_5$H$_2$]$^+$,
[Cr(H$_2$)$_6$]$^+$,
[VH$_2$O(H$_2$)$_3$]$^+$, and
[Cr(H$_2$)$_4$]$^+$.

\begin{table}
\begin{threeparttable} 
\caption{Mean absolute errors (MAE), mean signed errors (MSE), and maximum errors (MaxE) [kcal/mol] for dihydrogen complexes. CC refers to DLPNO-CCSD(T).}
\begin{tabular}{l r r r r r r r r}
\hline\hline 
            &	ph-AFQMC	&	CC	&	B3LYP	&	B97	&	M06	&	PBE0	&	$\omega$B97X-V	&	DSD-PBEP86	 \\ [0.5ex] 
\hline 
MAE  & 0.85 $\pm$ 0.21 & 1.82 & 1.43 & 0.93 & 2.50 & 0.75 & 1.43 & 1.09 \\
MSE  & 0.09 $\pm$ 0.21 & 1.75 & -1.36 & -0.67 & 1.94 & 0.33 & 1.08 & 1.04 \\
MaxE  & 1.51 $\pm$ 1.36 & 7.54 & 3.29 & 2.05 & 4.68 & 2.91 & 8.08 & 8.49 \\
\hline\hline
\end{tabular}
\label{table:Dihydrogentable}
\end{threeparttable}
\end{table} 

The relatively small system sizes of these dihydrogen complexes renders the ph-AFQMC calculations affordable even with the QZ basis set.  Therefore, for [Ni(H$_2$)$_4$]$^+$, which showed deviations $>$ 2 kcal/mol (see SI), we opted to do the full TZ/QZ extrapolation entirely within ph-AFQMC, and found better agreement.  In contrast, the scaling factor, i.e. the ratio between the correlation energies computed by ph-AFQMC and DLPNO-CCSD(T) at the TZ level was close to or more than 1.3 for [Co(H$_2$)$_6$]$^+$ and [Fe(H$_2$)$_6$]$^+$, a metric found in our previous work,\cite{shee2019achieving} so we also did TZ/QZ extrapolation entirely within ph-AFQMC in these cases, leading to good agreement. 
In the SI, we show that using ph-AFQMC/UHF to extrapolate gives similar results to the full treatment for the dihydrogen species. 


M06 yields the largest MAE (2.5 kcal/mol) while B97, PBE0, and ph-AFQMC have MAEs less than 1 kcal/mol.   
While ph-AFQMC and most density functionals (DFs) perform reasonably well for Cr(CO)$_5$H$_2$, especially given the relatively large experimental uncertainty, DSD-PBEP86 and DLPNO-CCSD(T) are off by 6-8 kcal/mol.  We note that in the next section DSD-PBEP86 is seen to over-stabilize all carbonyl complexes. 
$\omega$B97X-V drastically overestimates the BDE of the [Ni(H$_2$)$_4$]$^+$ complex, with a deviation of 8.08 kcal/mol.  Indeed, as will be shown, this functional over-stabilizes all Ni complexes.

\subsection{Aqua Complexes}
\begin{figure}[H]
    \centering
    \includegraphics[width=14cm]{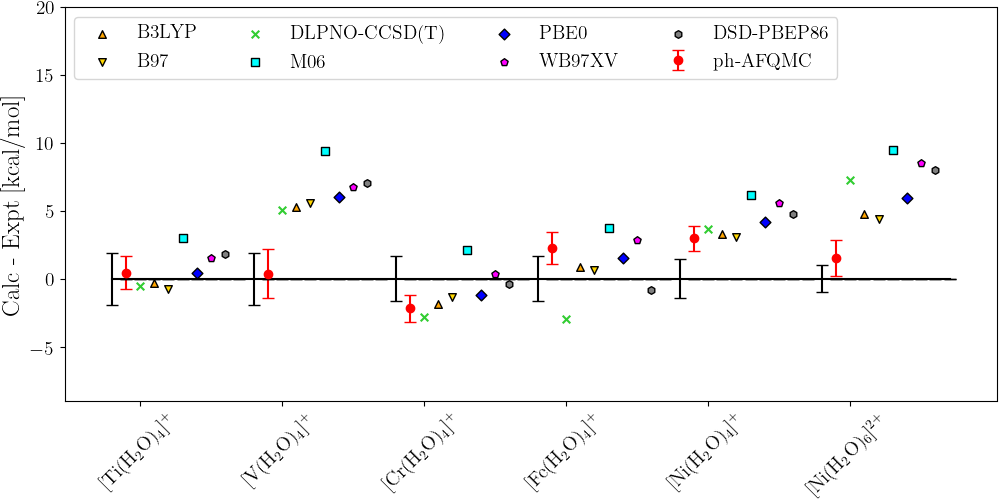} 
    \caption{Deviations [kcal/mol] of computational methods for the aqua set of bond dissociation reactions where the H$_2$O that leaves is given at the end of the formula.} 
    \label{fig:LigandPlot_H2O}
\end{figure}

As shown in Fig. \ref{fig:LigandPlot_H2O} and Table \ref{table:Aquatable}, ph-AFQMC also yields accurate results for the hexaaqua complex [Ni(H$_2$O)$_6$]$^{2+}$ and the tetraaqua complexes
[Cr(H$_2$O)$_4$]$^+$,
[Ni(H$_2$O)$_4$]$^+$,
[Ti(H$_2$O)$_4$]$^+$,
[V(H$_2$O)$_4$]$^+$,
and	[Fe(H$_2$O)$_4$]$^+$.  While all other methods seem to overbind these complexes, as can be seen by large and positive MSEs, ph-AFQMC appears to predict the BDEs in a relatively balanced manner.

\begin{table}
\begin{threeparttable} 
\caption{Mean absolute errors (MAE), mean signed errors (MSE), and maximum errors (MaxE) [kcal/mol] for aqua complexes. CC refers to DLPNO-CCSD(T).}
\begin{tabular}{l r r r r r r r r}
\hline\hline 
            &	ph-AFQMC	&	CC	&	B3LYP	&	B97	&	M06	&	PBE0	&	$\omega$B97X-V	&	DSD-PBEP86	 \\ [0.5ex] 
\hline 
MAE  & 1.61 $\pm$ 0.84 & 3.70 & 2.72 & 2.61 & 5.65 & 3.20 & 4.25 & 3.81 \\
MSE  & 0.89 $\pm$ 0.84 & 1.60 & 1.99 & 1.91 & 5.65 & 2.81 & 4.25 & 3.40 \\
MaxE  & 2.96 $\pm$ 1.71 & 7.24 & 5.26 & 5.54 & 9.49 & 5.98 & 8.48 & 7.99 \\
\hline\hline
\end{tabular}
\label{table:Aquatable}
\end{threeparttable}
\end{table} 

In the case of [Ni(H$_2$O)$_6$]$^{2+}$, the scaling factor was below 0.6, which indicates a poor match between the correlation energies of ph-AFQMC and DLPNO-CCSD(T). As full TZ/QZ extrapolation within ph-AFQMC was unaffordable in the present version of our code implementation due to prohibitively high required device memory, we opted to do the extrapolation with a single-determinant (UHF) trial based QMC in place of DLPNO-CCSD(T) and found good results.
Notably, all other methods overestimate the BDE for this molecule by at least 5 kcal/mol, well outside the reported experimental uncertainty.  All DFs and DLPNO-CCSD(T) give errors in excess of 5 kcal/mol for this molecule.  Similarly, we 
performed the extrapolation with ph-AFQMC/UHF for [V(H$_2$O)$_4$]$^+$, on the basis of disagreement of experiment rather than the scaling factor, and found that the deviation went from 4.03 $\pm$ 1.95 kcal/mol with the DLPNO-CCSD(T) extrapolation to 0.35 $\pm$ 2.63 kcal/mol with the ph-AFQMC/UHF extrapolation. The other methods have errors around 5-9 kcal/mol for this molecule.  These findings suggest that these two species exhibit significant multireference character. 

On average, as seen in Table \ref{table:Aquatable}, the accuracy of CC and DFT methods for metal-aqua complexes is similar with MAE's between 2.61 (B97) and 5.65 (M06) kcal/mol.  
The MAE of ph-AFQMC is 1.61 $\pm$ 0.84 kcal/mol, 
with a MaxE of 2.96 $\pm$ 1.71 kcal/mol 
found for
the [Ni(H$_2$O)$_4$]$^+$ species. 
We note that all methods overestimate the BDE of this molecule, although not by a huge amount, especially in light of the experimental error bars.  It is thus possible that the experimental value for this case should be reinvestigated.

\subsection{Ammonia Complexes}
\begin{figure}[H]
    \centering
    \includegraphics[width=14cm]{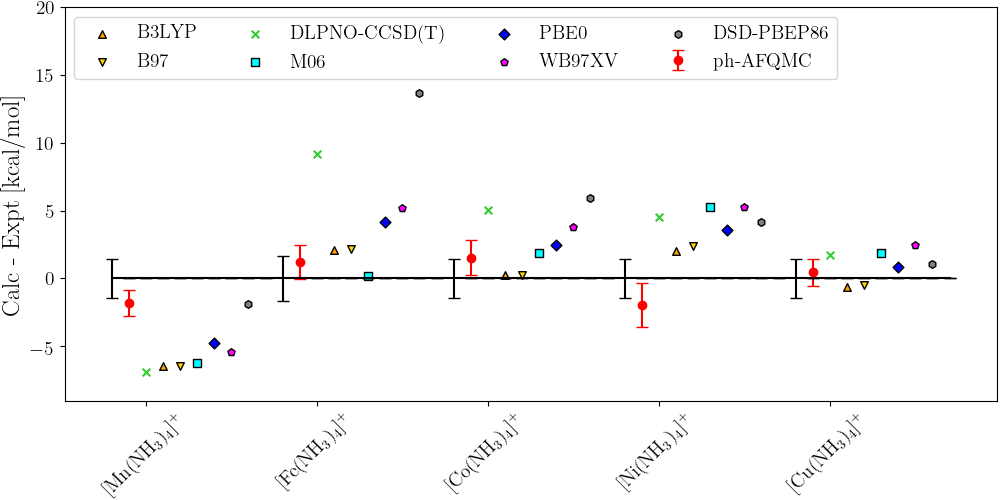} 
    \caption{Deviations [kcal/mol] of computational methods for the aqua set of bond dissociation reactions where the NH$_3$ that leaves is given at the end of the formula.} 
    \label{fig:LigandPlot_NH3}
\end{figure}

Fig. \ref{fig:LigandPlot_NH3} and Table \ref{table:Ammoniatable} summarize the performance of the computational methods for the tetraammonia complexes: [Co(NH$_3$)$_4$]$^+$,	
[Ni(NH$_3$)$_4$]$^+$,
[Mn(NH$_3$)$_4$]$^+$,
[Cu(NH$_3$)$_4$]$^+$, and
[Fe(NH$_3$)$_4$]$^+$.

\begin{table}
\begin{threeparttable} 
\caption{Mean absolute errors (MAE), mean signed errors (MSE), and maximum errors (MaxE) [kcal/mol] for ammonia complexes. CC refers to DLPNO-CCSD(T).}
\begin{tabular}{l r r r r r r r r}
\hline\hline 
            &	ph-AFQMC	&	CC	&	B3LYP	&	B97	&	M06	&	PBE0	&	$\omega$B97X-V	&	DSD-PBEP86	 \\ [0.5ex] 
\hline 
MAE  & 1.39 $\pm$ 0.87 & 5.46 & 2.29 & 2.36 & 3.09 & 3.15 & 4.44 & 5.36 \\
MSE  & -0.12 $\pm$ 0.87 & 2.71 & -0.55 & -0.42 & 0.60 & 1.25 & 2.26 & 4.61 \\
MaxE  & 1.95 $\pm$ 2.16 & 9.15 & 6.48 & 6.45 & 6.22 & 4.74 & 5.45 & 13.69 \\
\hline\hline
\end{tabular}
\label{table:Ammoniatable}
\end{threeparttable}
\end{table} 

[Mn(NH$_3$)$_4$]$^+$ is a difficult case for all methods. 
DSD-PBEP86 and ph-AFQMC, with deviations of $\sim$2 kcal/mol, performed better compared to other methods which showed errors of $\sim$6 kcal/mol. This reaction involves the only 2 molecules (i.e. [Mn(NH$_3$)$_4$]$^+$ and [Mn(NH$_3$)$_3$]$^+$) 
where we had to run separate ph-AFQMC calculations with population control
 because the imaginary trajectories were not convincingly equilibrated by 15 $\beta$.  Additionally, there were many CAS convergence issues that prevented us from running larger CASSCF active spaces to check the convergence. Further investigation will be required. DLPNO-CCSD(T) and the remaining DFs perform particularly poorly for this molecule with errors around or above 5 kcal/mol.  

We note that [Ni(NH$_3$)$_4$]$^+$ is another case for which basis set extrapolation with ph-AFQMC/UHF 
reduced
the deviation from experiment.  As before, this may indicate multireference character, which causes all other methods to significantly overbind the ammonia ligand. 

Overall, ph-AFQMC, B3LYP, B97, and M06 have notably small MSEs.  ph-AFQMC is outstanding here with respect to MAE (1.39 $\pm$ 0.87 kcal/mol) and MaxE (1.95 $\pm$ 2.16 kcal/mol) while other methods show a MaxE around 6-14 kcal/mol for these complexes.  DLPNO-CCSD(T) and  DSD-PBEP86 showed the largest deviations with MAEs of 5.46 and 5.36 kcal/mol, respectively. They show extreme errors for [Fe(NH$_3$)$_4$]$^+$ in particular, with MaxEs of 9-14 kcal/mol.

\subsection{Carbonyl Complexes}
\begin{figure}[H]
    \centering
    \includegraphics[width=14cm]{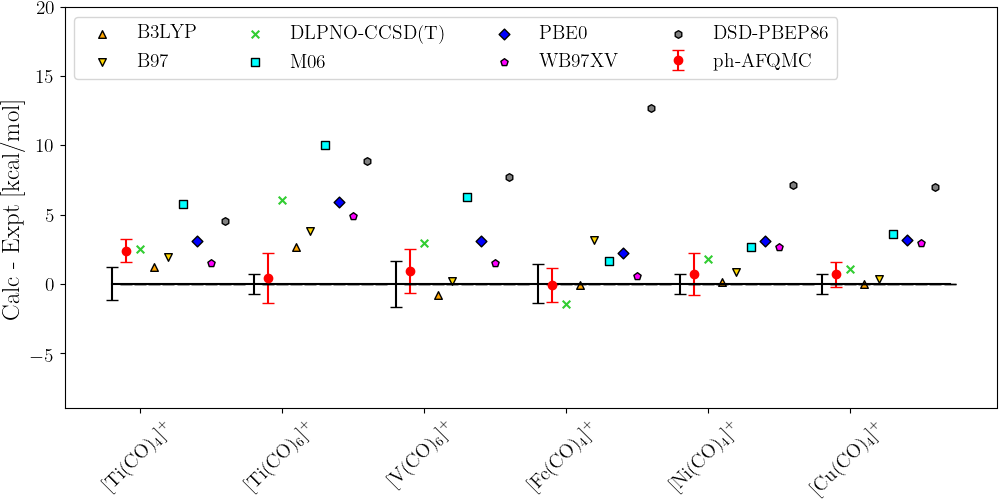} 
    \caption{Deviations [kcal/mol] of computational methods for the carbonyl set of bond dissociation reactions where the CO that leaves is given at the end of the formula.} 
    \label{fig:LigandPlot_CO}
\end{figure}

As shown in Fig. \ref{fig:LigandPlot_CO} and Table \ref{table:Carbonyltable}, ph-AFQMC also performed well for the species with all carbonyl ligands: [Ti(CO)$_6$]$^+$,	
[Ni(CO)$_4$]$^+$,
[Cu(CO)$_4$]$^+$,	
[Ti(CO)$_4$]$^+$,
[Fe(CO)$_4$]$^+$, and
[V(CO)$_6$]$^+$.	
In particular, ph-AFQMC is the only method to predict a BDE 
close to
the experimental value for [Ti(CO)$_6$]$^+$ (although B3LYP is just outside the AFQMC
statistical error bars). 

\begin{table}
\begin{threeparttable} 
\caption{Mean absolute errors (MAE), mean signed errors (MSE), and maximum errors (MaxE) [kcal/mol] for carbonyl complexes. CC refers to DLPNO-CCSD(T).}
\begin{tabular}{l r r r r r r r r}
\hline\hline 
            &	ph-AFQMC	&	CC	&	B3LYP	&	B97	&	M06	&	PBE0	&	$\omega$B97X-V	&	DSD-PBEP86	 \\ [0.5ex] 
\hline 
MAE  & 0.87 $\pm$ 0.72 & 2.65 & 0.83 & 1.71 & 4.99 & 3.43 & 2.35 & 7.99 \\
MSE  & 0.85 $\pm$ 0.72 & 2.18 & 0.52 & 1.71 & 4.99 & 3.43 & 2.35 & 7.99 \\
MaxE  & 2.39 $\pm$ 1.46 & 6.07 & 2.64 & 3.80 & 10.02 & 5.90 & 4.88 & 12.68 \\
\hline\hline
\end{tabular}
\label{table:Carbonyltable}
\end{threeparttable}
\end{table} 

DSD-PBEP86 gives an extremely large deviation of 12.68 kcal/mol for [Fe(CO)$_4$]$^+$, and in fact overpredicts all carbonyl species in this set,
with an MAE and MSE of $\sim$ 7.99 kcal/mol.  M06 has the second largest MAE (4.99 kcal/mol) and MaxE (10.02 kcal/mol for [Ti(CO)$_6$]$^+$) among all methods.  
For these carbonyl complexes, both ph-AFQMC and B3LYP showed outstanding performance with balanced predictions (low MSEs), MAEs of $<$ 1 kcal/mol, and MaxEs  of $\sim$ 2.5 kcal/mol.  

In the case of [Ti(CO)$_4$]$^+$, all methods predict BDEs above the experimental measurement.  We therefore suggest, for a future study, that the experimental value be examined carefully.

\subsection{Miscellaneous Complexes}
\begin{figure}[H]
    \centering
    \includegraphics[width=14cm]{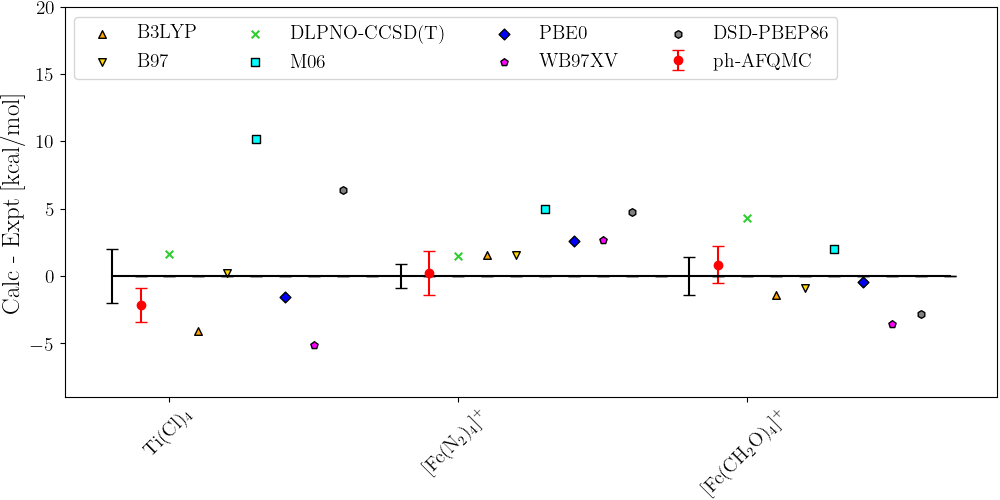} 
    \caption{Deviations [kcal/mol] of computational methods for the other reactions where the ligand that leaves is given at the end of the formula.} 
    \label{fig:LigandPlot_MISC}
\end{figure}

As can be seen in Fig. \ref{fig:LigandPlot_MISC}, ph-AFQMC continues to predict  consistently accurate BDEs for these three complexes.  While 
a statistical analysis of three compounds is likely not rigorously meaningful, we nonetheless provide a summary in Table \ref{table:Miscellaneoustable}, for completeness.

\begin{table}
\begin{threeparttable} 
\caption{Mean absolute errors (MAE), mean signed errors (MSE), and maximum errors (MaxE) [kcal/mol] for miscellaneous complexes. CC refers to DLPNO-CCSD(T).}
\begin{tabular}{l r r r r r r r r}
\hline\hline 
            &	ph-AFQMC	&	CC	&	B3LYP	&	B97	&	M06	&	PBE0	&	$\omega$B97X-V	&	DSD-PBEP86	 \\ [0.5ex] 
\hline 
MAE  & 1.07 $\pm$ 1.19 & 2.45 & 2.37 & 0.89 & 5.72 & 1.56 & 3.80 & 4.66 \\
MSE  & -0.37 $\pm$ 1.19 & 2.45 & -1.35 & 0.27 & 5.72 & 0.17 & -2.01 & 2.76 \\
MaxE  & 2.16 $\pm$ 2.36 & 4.29 & 4.12 & 1.54 & 10.18 & 2.59 & 5.15 & 6.37 \\
\hline\hline
\end{tabular}
\label{table:Miscellaneoustable}
\end{threeparttable}
\end{table} 

The experimental uncertainty corresponding to the measured Ti(Cl)$_4$ BDE is the highest among the molecules included in this study, at 2 kcal/mol. Most of the methods give reasonable performance except DSD-PBEP86, M06  and $\omega$B97X-V.  The first two overestimated the BDE by $\sim$ 6-10 kcal/mol while the latter underestimated it by 5.15 kcal/mol. 

We note that all DFT methods overestimate the BDE of [Fe(N$_2$)$_4$]$^+$, with M06 and DSD-PBEP86 yielding deviations of around 5 kcal/mol.	

The formaldehyde ligands make [Fe(CH$_2$O)$_4$]$^+$ the largest molecule studied in this work. $\omega$B97X-V and DSD-PBEP86 yield deviations of $\sim$ -3 kcal/mol while DLPNO-CCSD(T) yields of a deviation around $\sim$ 3 kcal/mol. 


\subsection{Performance for the Entire Test Set}

The statistical performance of each computational method over all ligand types is summarized in Table \ref{table:MAEtable}.  We note that the average experimental uncertainty is 1.03 kcal/mol.  

\begin{table}[H]
\captionsetup{margin=2cm}
\centering
\caption{Mean absolute errors (MAE), mean signed errors (MSE), and maximum errors (MaxE) [kcal/mol] of ph-AFQMC, DLPNO-CCSD(T), and DFT results and other methods for the 34 molecule subset shown in Fig. \ref{fig:coordination_compounds_final}. The values are sorted by MAE. 
The ph-AFQMC deviations incorporate both the experimental uncertainty and the statistical uncertainty.}  
\begin{tabular}{l r r r}
\hline\hline 
            & MAE & MSE & MaxE    \\ [0.5ex] 
\hline 
ph-AFQMC         & 1.09 $\pm$ 0.27 & 0.30 $\pm$ 0.27 & 2.96 $\pm$ 1.71  \\
B97      		 & 1.57 & 0.33 & -6.45 \\
B3LYP            & 1.76 & -0.32 & -6.48 \\
PBE0             & 2.08 & 1.43 & 5.98  \\
$\omega$B97X-V   & 2.74 & 1.77 & 8.48  \\
DLPNO-CCSD(T)    & 2.89 & 2.00 & 9.15  \\
DSD-PBEP86/2013  & 3.73 & 3.36 & 13.69 \\
M06              & 3.87 & 3.27 & 10.18  \\
\hline\hline
\end{tabular}
\label{table:MAEtable}
\end{table} 

ph-AFQMC, B97, and B3LYP have near-zero MSEs, while all other methods systematically overestimate the BDEs.  ph-AFQMC outperforms all DFT functionals and DLPNO-CCSD(T), with an MAE of  1.09 $\pm$ 0.27 kcal/mol and MaxE of 2.96 $\pm$ 1.71 kcal/mol.
DLPNO-CCSD(T) performs worse than most of the hybrid functionals in the study, with MAE and MaxE of 2.89 and 9.15 kcal/mol, respectively.  In light of the average uncertainty in the experimental measurements reported above, the B97 and B3LYP functionals arguably yield, on average, comparable accuracy to ph-AFQMC, with MAEs of 1.57 and 1.76 kcal/mol, respectively.
Yet the MaxE's of 6.45 and 6.48 kcal/mol are more than twice as large as that from ph-AFQMC, and would be considered much too large for many predictive applications. $\omega$B97X-V achieved a similar accuracy as DLPNO-CCSD(T), with MAE and MaxE of 2.74 and 8.48 kcal/mol, respectively.  This performance is rather satisfactory given that there were no transition metals in the training set used to fit the 10 empirical parameters in the functional.\cite{mardirossian2014omegab97xv}  In contrast, the Minnesota functional, M06, is heavily parameterized and results in the largest MAE of 3.87 kcal/mol.  The poor performance of M06 for transition-metal complexes was also mentioned in our group's previous paper\cite{coskun2016evaluation} and in the work of Grimme and co-workers\cite{steinmetz2013benchmark}.  In contrast to the high accuracy achieved by double-hybrid functionals for organic molecules \cite{goerigk2017look,santra2019minimally}, the DSD-PBEP86 functional for this dataset yielded an MAE of 3.73 kcal/mol and MaxE of 13.69 kcal/mol.

According to Grimme and co-workers, DFs with a smaller amount of HF exchange tend to perform better than those with larger percentages.\cite{steinmetz2013benchmark}  We see a similar trend that B97 (19.43\%  HF exchange) gives the best performance for this dataset while M06 (27\% HF exchange) and DSD-PBEP86 ($\sim$ 70\% HF exchange) perform the worst. PBE0 with an MAE of 2.17 kcal/mol is slightly worse than B3LYP and B97; however, it yields good results for dihydrogen complexes.  

We attempted to correlate a number of multireference diagnostics, such as the fractional occupation number weighted electron density (FOD) \cite{grimme2015practicable,bauer2017fractional} and 
the square of the leading CI coefficient in the CASSCF calculation,\cite{momeni2015td}
with errors from DLPNO-CCSD(T).  However, no significant correlation was found.  This is consistent with previous studies reporting similar inefficacy for transition metal systems.\cite{Xu_Truhlar_JCTC_2015,cheng2017bond,fang2017prediction,aoto2017arrive,momeni2015td}  We emphasize the need for further investigation and development of multireference diagnostics that can reliably identify the presence of strong correlation effects and thus signal caution to users of single-reference methods such as DFT and CCSD(T).  One promising approach involves examining the deviation of $\langle S ^2_{UHF} \rangle$ from spin-pure values, in conjunction with the use of an orbital-optimized method, e.g. MP$n$, to rule out artificial symmetry breaking\cite{lee2019distinguishing}. 

For reactions involving Sc, Ti, V, and Cr centers, our ph-AFQMC results are typically in good agreement with experiment even when relatively small active spaces are employed in the trial wavefunction.  Such calculations need only use the MINAO basis set to specify the 3$d$ orbitals as inputs for the AVAS procedure for selecting the active space. For the remaining metals, larger active spaces (i.e. including higher-lying virtual orbitals) are required, and we therefore used the ANO-RCC basis for AVAS, specifying both the 3$d$ and 4$d$ atomic orbitals to account for the double-shell effect.\cite{bauschlicher1988atomic,andersson1992excitation,shee2018gpu}

As a number of functionals were trained utilizing larger basis sets than the one employed in this work, we note that the results may change slightly if such optimal basis sets had been employed.  We did investigate the basis set dependence for the double-hybrid functional, as the MP2-like part is known to perform better with a basis larger than TZ to more closely approach the complete basis set limit.\cite{goerigk2010efficient,chan2015accurate}
We found for the largest outliers for DSD-PBEP86 that using a QZ basis set for the single-point energy calculations did not significantly change the results.  For example, the calculated BDEs of [Fe(NH$_3$)$_4$]$^+$ in TZ and QZ deviate from experiment by 13.69 and 13.89 kcal/mol, respectively.  

\subsection{Discussion}

The results we have obtained lead to interesting observations concerning all three classes of 
approaches considered in this paper: AFQMC, DLPNO-CCSD(T), and DFT.  These observations have implications that go beyond the current data set. Our previous AFQMC study on transition metal containing dimers\cite{shee2019achieving} could be viewed as addressing a very special subset of unusual and difficult molecules from an electronic structure point of view.
In particular, these systems are
coordinatively unsaturated, with nearly degenerate electronic states in a number of cases, and of a form rarely present in important chemical systems relevant to practical applications in biology and materials science.  In contrast, the present data set contains many typical bonding motifs, namely
four and six coordinated metal-ligand complexes, 
although the oxidation states are lower than is usually found in 
condensed phase systems.
  Arguably, a system such as the water splitting complex in Photosystem II poses a much more difficult quantum chemistry problem than the molecules considered here.  A method that displays a significant number of outliers in our present data set would be difficult to trust as reliable if applied to a strongly interacting, multi-metal complex with a large number of low lying electronic states.

The AFQMC results satisfy all of the criteria one could reasonably expect (given the uncertainties in the experimental data) for true benchmark performance.  The largest deviation from experiment is less than 3 kcal/mol, often cited as the target for “transition metal chemical accuracy”,\cite{deyonker2007quantitative} and close to being within the cited experimental error bars.  For most of the ligands studied, the maximum deviation is closer to 2 kcal/mol and well within experimental error.  Results reliably improve (sometimes considerably) as the quality of the calculation is increased, e.g. via an upgrade in the basis set extrapolation method. In fact, the error for the [Ni(H$_2$O)$_4$]$^+$ molecule, which represents the MaxE of ph-AFQMC in Table \ref{table:MAEtable}, can be reduced to less than 1 kcal/mol when utilizing QMC/UHF rather than DLPNO-CCSD(T) for the basis set extrapolation 
(we indicate in Tables S4 and S5 that extrapolating with QMC/UHF will produce equally good if not better final BDEs for a representative selection of molecules, suggesting that such extrapolation is to be preferred, if computationally feasible, in future studies).  With this update the MaxE of ph-AFQMC would be lowered to 2.39  $\pm$ 1.46 kcal/mol, for [Ti(CO)$_4$]$^+$, which is a rather outstanding result in light of the experimental uncertainty.  The overall mean unsigned deviation from experiment of 1.1 kcal/mol is highly satisfactory.  It is in fact not obvious how much of this deviation is due to errors in the theory and how much to errors in the experiment.  In our transition metal dimer publication, it is noteworthy that when new (and more reliable) experiments were released after the calculations were completed (but prior to publication), agreement of AFQMC with these results was significantly better than with older values. In the absence of significantly more accurate experiments, it is hard to imagine a better performance from a tractable theoretical approach.

The DLPNO-CCSD(T) results, in contrast, reveal a large number of major outliers (with a maximum outlier of 9.15 kcal/mol) across every single ligand series (maximum deviations for the individual series range from 4.29 kcal/mol to 9.15 kcal/mol). The DLPNO approximations are likely \emph{not} the most significant sources of error, given that we use the tightest possible cutoff parameters, and in light of the results in Ref. \citenum{liakos2015exploring}.  In addition, due to the relatively small size of the dissociating ligand, it is reasonable to expect some degree of cancellation in the localization errors.  It is most likely that excitations of  higher order than (T) are required
for consistently high accuracy,  
though we note that it would be a useful future investigation to probe the effects of utilizing orbitals from, e.g., an unrestricted DF calculation.
Regardless of the source of the errors, the implication is that much more expensive (and poorly scaling) variants of coupled cluster will be needed to converge this approach to chemical accuracy for transition metal containing systems.  Now that benchmark values are available (via our AFQMC results) for both transition metal containing dimers and small four and six coordinated complexes (comprising roughly 80 systems in all), we look forward to 
alternative CC approximations being rigorously evaluated using this data. At that point, assuming that comparable benchmark quality can be achieved, it will be interesting to compare the computational requirements, and scaling with system size, of both methods.

The DFT results shown here are far from a comprehensive survey of the various flavors of functionals currently available, but do contain a number of qualitatively different functionals as well as several of the most widely used approaches. A striking observation is that the three best performing functionals- by a considerable margin-  were published more than 20 years ago.  Despite the use of considerably more sophisticated functional forms, the performance of the three more recent functionals (wB97X-V, DSD-PBDP86, and M06) have substantially worse average errors, and larger and more frequent outliers, than the older approaches. It should also be noted that the best performing DFT approaches work substantially better than DLPNO-CCSD(T). This observation is in accordance with the proposition put forth along these lines by Truhlar and coworkers several years ago, which has been the subject of considerable controversy in the literature.\cite{Xu_Truhlar_JCTC_2015,aoto2017arrive,shee2019achieving}  While 
 one could ultimately 
converge
 coupled cluster--based methods to a benchmark level of accuracy by including higher (and considerably more expensive) levels of theory, what is going to be necessary and sufficient to accomplish that convergence is apparently more demanding than some of the earlier papers in this debate have suggested. 

Our results cast doubt as to whether the newer DFT models use a functional form that is an actual improvement from the point of view of transition metal chemistry, as the incorporation of asymptotically correct exchange, non-local correlation, MP2 contributions, kinetic energy density-dependence and/or a greater number of parameters appears not to yield improved accuracy over simpler hybrid GGA forms.
As in the case of 
typical
machine learning problems, consideration of additional parameters generally leads to better performance when the test cases are similar to the molecules in the training set, i.e. when 
direct
interpolation is performed. Extrapolation outside of the training set, however, is a very different proposition. The lack of confidence in the experimental values for transition metal energetics has deterred extensive incorporation of data of the type we have studied here into the process of fitting DFT functionals.  Our benchmark level of agreement with experiment should enable new efforts, incorporating the data we have validated here, to proceed with more confidence.  And it is of course possible that one of the many DFT functionals that we have not tested in this paper would improve upon any of the results presented above.  Again, data is now available to rigorously interrogate such a proposition.

The performance of the two best performing methods, B3LYP and B97, is quite remarkable considering their vintage and relatively small number of fitting parameters (3 and 10, respectively).  It is interesting that whereas B97 was clearly superior for the transition metal dimer data set, the results for the present data set are much closer in average and maximum error.  For calculations 
of
large, 
transition
metal--containing systems, we would view either of these alternatives as the best currently available, particularly given the extensive experience with them over the past several decades (although not of benchmark quality, in view of the presence of a significant number of outliers in the 3-7 kcal/mol error range).  If the AFQMC calculations can be scaled up to address systems with 50-100 atoms, perhaps
by using localized orbital techniques, a combination of AFQMC benchmarks followed by B97 or B3LYP modeling of a larger set of conformations (including environmental effects such as solvation), could provide a path towards calculations 
of high enough quality
to understand reaction mechanisms, identify intermediates, and contribute to molecular design efforts. 

\section{Conclusions}
Our ph-AFQMC 
approach
has produced reliable theoretical values for BDEs in 3$d$ transition metal coordination complexes.
Our results demonstrate that future, predictive benchmarking should employ CAS trial
wavefunctions
in the TZ basis with QMC/UHF for CBS extrapolation. The MAEs of the DFs considered in this study are in general quite satisfactory, but the occasional presence of large, unsystematic errors leaves cause for concern. The performance of methods by MAE from best to worst is ph-AFQMC, B97, B3LYP, PBE0 DLPNO-CCSD(T), $\omega$B97X-V, DSD-PBEP86, and M06, respectively.

We envision that this dataset of gas-phase BDEs may prove useful for the development of new approximate methods, 
and
new DFs.  The reliability of the ph-AFQMC method, namely its ability to compute accurate gas-phase energetics in a reasonable amount of wall-time, will enable the development of accurate force-fields for metal ion interactions with various ligands.  The method will also help in a forthcoming investigation of DFT's ability to predict solution-phase properties.  For instance, we are now in a position to answer the question:  are 
errors found in recent studies of
aqueous pK$_a$'s\cite{Jerome_Hughes_Friesner_2016} and redox potentials\cite{coskun2016evaluation}
due inherently to deficiencies in the quantum-chemical electronic structure description or in the implicit solvent models employed, or both?

For the systems in this work, we were generally able to converge the BDEs with respect to active space size of the trial wavefunctions.  However, moving on to larger systems, perhaps containing multiple metals or bulky ligands, we anticipate that the relevant active space sizes will overcome conventional CASSCF algorithms and available computing resources.  Investigations along these lines are currently underway, as are efforts to implement a localized orbital approach to ph-AFQMC.  

\begin{acknowledgement}
D.R.R. acknowledges funding from NSF CHE-1839464. S.Z. acknowledges funding from DOE DE-SC0001303. 
This research used resources of the Oak Ridge Leadership Computing Facility at the Oak Ridge National Laboratory, which is supported by the Office of Science of the U.S. Department of Energy under Contract No. DE-AC05-00OR22725.
This work used the Extreme Science and Engineering Discovery Environment (XSEDE), which is supported by National Science Foundation grant number ACI-1548562. In particular, we used San Diego Computing Center's Comet resources under grant number TG-CHE190007 and allocation ID COL151.
The Flatiron Institute is a division of the Simons Foundation.
We would like to thank 
Elvira Sayfutyarova for helpful discussions regarding AVAS.

\end{acknowledgement}

\bibliography{References}

\makeatletter
\setlength\acs@tocentry@width{4.45cm}
\setlength\acs@tocentry@height{8.25cm}
\makeatother

\begin{tocentry} 
\includegraphics[width=6.25cm]{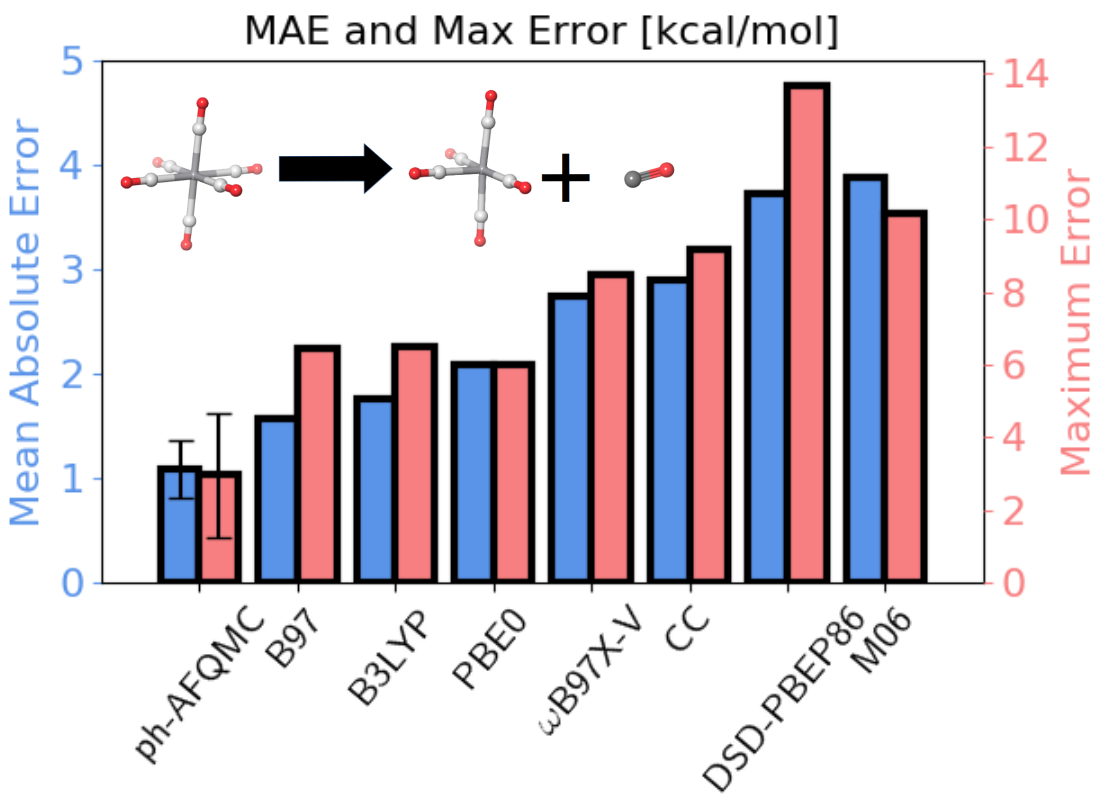}
\end{tocentry}

\end{document}